\newcommand{\rem}[1]{}
\newcommand{\eqref}[1]{(\ref{#1})}
\documentclass[12pt]{iopart}
\usepackage{hyperref}
\usepackage{epsfig}

\begin{document}

\title{Solitons and exact velocity quantization of incommensurate sliders}

\author{Nicola Manini$^{1,2}$, Marco Cesaratto$^{1,2}$,
Giuseppe E. Santoro$^{2,3,4}$, Erio Tosatti$^{2,3,4}$,
and Andrea Vanossi$^{5}$}
\address{$^1$
Dipartimento di Fisica and CNR-INFM, Universit\`a di Milano,
Via Celoria 16, 20133 Milano, Italy}
\address{$^2$
International School for Advanced Studies (SISSA), Via Beirut 2-4,
I-34014 Trieste, Italy}
\address{$^3$ INFM Democritos National Simulation Center}
\address{$^4$ International Centre for Theoretical Physics
(ICTP), P.O.Box 586, I-34014 Trieste, Italy}
\address{$^5$ CNR-INFM National Research Center S3 and Department
of Physics, \\University of Modena and Reggio Emilia, Via Campi
213/A, 41100 Modena, Italy}

\date{\today}

\begin{abstract}
We analyze in some detail the recently discovered velocity quantization
phenomena in the classical motion of an idealized one-dimensional solid
lubricant, consisting of a harmonic chain interposed between two periodic
sliders.
The ratio $w = v_{\rm cm}/v_{\rm ext}$ of the chain center-of-mass velocity
to the externally imposed relative velocity of the sliders is pinned to
exact ``plateau'' values for wide ranges of parameters, such as sliders
corrugation amplitudes, external velocity, chain stiffness and dissipation,
and is strictly determined by the commensurability ratios alone.
The phenomenon is caused by one slider rigidly dragging
the density solitons (kinks/antikinks)
that the chain forms with the other slider.
Possible consequences of these results for some real systems are discussed.
\end{abstract}


\maketitle

\section{Introduction to the model}

\begin{figure}
\centerline{
(a)\epsfig{file=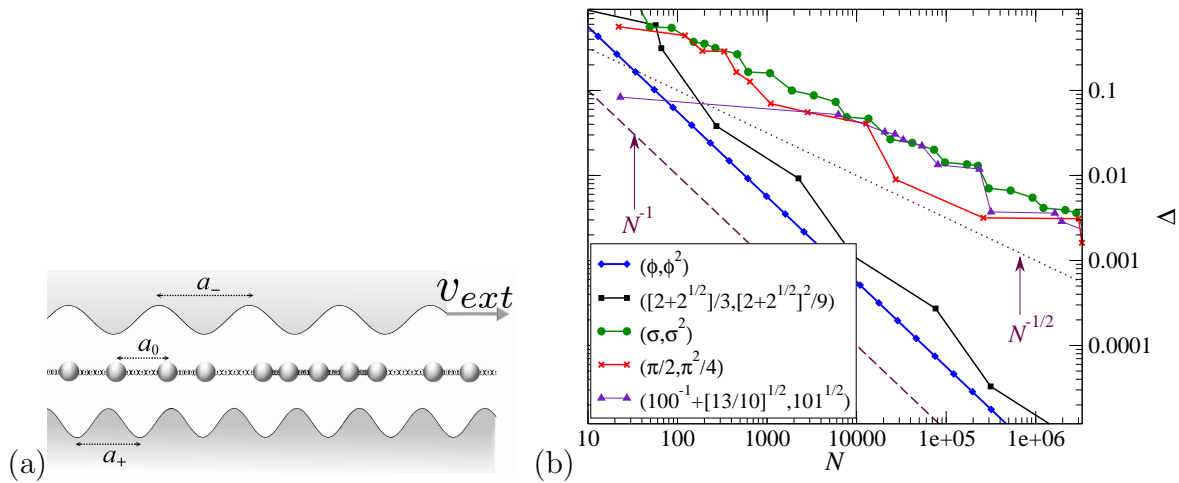,height=27mm,angle=0,clip=}
\hfill
(b)\epsfig{file=delta.eps,height=62mm,angle=0,clip=}}
\caption{\label{model:fig}
(a) A sketch of the model with the two periodic sliders (periods $a_+$ and
$a_-$) and the lubricant harmonic chain of rest nearest-neighbor length
$a_0$.
(b) The finite-size phase error $\Delta$ of Eq.~\eqref{Delta:eq} (only the
$N$ values where $\Delta$ is smaller than for any other $N'<N$ are shown)
decreases as a function of size $N$.
Note that the error drops fast as $N^{-1}$ for golden mean and other
quadratic incommensurabilities $(r_+,r_-=r_+^2)$, but slowly as $N^{-1/2}$
for spiral mean and other non-quadratic incommensurabilities.
}
\end{figure}

\begin{table}
\begin{center}
\begin {tabular}{ll}
\hline
energy & $a_+ \, F_+$ \\
velocity $v$&  $a_+^{1/2}\, F_+^{1/2} \, m^{-1/2}$ \\
time & $a_+^{1/2}\, F_+^{-1/2} \, m^{1/2}$ \\
spring  constant $K$ & $a_+^{-1} \, F_+$ \\
viscous friction $\gamma$ & $a_+^{-1/2}\, F_+^{1/2} \, m^{1/2}$ \\
\hline
\end{tabular}
\end{center}
\caption{\label{units:tab}
Natural units for several physical quantities in a system where length,
force and mass are measured in units of $a_+$, $F_+$, and $m$ respectively.
}
\end{table}

The sliding friction of hard incommensurate crystals has
stimulated for over a decade the study of non-linear classical
dynamical systems -- belonging to the class of generalized
Frenkel-Kontorova (FK) models \cite{BraunBook} -- which can show a
rich variety of dynamical behavior, ranging from regular to
chaotic motion \cite{Kapitaniak}.
In particular, in presence of a uniform external driving,
the length-scale competition between different spatial
periodicities can give rise to intriguing frequency
incommensurations and phase locking phenomena.
In this paper we discuss and analyze in depth the striking exact
velocity quantization phenomena recently reported in
a one-dimensional (1D) non-linear model inspired by the tribological
problem of two sliding surfaces with a thin solid lubricant layer in
between \cite{Vanossi06,Santoro06}.
The model ``layer'' consists here of a chain of $N$ harmonically
interacting particles interposed between two rigid generally (but
not necessarily) incommensurate sinusoidal substrates representing
the two ``sliding crystals'', sketched in Fig.~\ref{model:fig}a,
externally driven at a constant relative velocity $v_{\rm ext}$.
The equation of motion of the $i$-th particle is:
\begin{eqnarray} \label{eqmotion:eqn}
m\ddot{x}_i &=&  -\frac{1}{2} \left[ F_+ \sin{k_+ (x_i-v_+t)}
+ F_- \sin{k_-(x_i-v_-t)}\right] 
\nonumber \\
&+& K (x_{i+1}+x_{i-1}-2x_i) - \gamma \sum_{\pm} (\dot{x}_i - v_{\pm}) \;,
\end{eqnarray}
where $m$ is the mass of the $N$ particles, $K$ is the chain spring
constant, and $k_{\pm}=(2\pi)/a_{\pm}$ are the wave-vector periodicities of
potentials representing the two sliders, moving at velocities $v_{\pm}$.
We choose the reference frame where $v_+$ = 0, so that $v_{\rm ext} \equiv
v_- - v_+=v_-$.
$\gamma$ is a phenomenological parameter substituting for various sources
of dissipation. Dissipation is required to achieve a stationary state, but
has otherwise no major role in the following.
$F_{\pm}$ are the amplitudes of the forces representing the sinusoidal
corrugation of the two sliders (we will commonly use $F_-/F_+=1$ but we
did check that our results are more general).
We take $a_+=1$, $m=1$, and $F_+=1$ as our basic units, so that the natural
units for all quantities used to characterize the model are listed in
Table~\ref{units:tab}.

The relevant length ratios \cite{vanErp99,Vanossi00} are therefore
$r_{\pm}=a_{\pm}/a_0$; we take, without loss of generality, $r_->r_+$.
The inter-particle equilibrium length $a_0$, not entering explicitly the
equation of motion (\ref{eqmotion:eqn}), appears via the boundary
conditions, which are taken to be periodic (PBC), $x_{N+1}=x_1+N \, a_0$,
to enforce a fixed-density condition for the chain \cite{BraunBook}, with a
{\em coverage} $r_+$ of chain atoms on the ``denser'' substrate of
periodicity $a_+$.
Technically, the PBCs can be realized in two different manners.
The first (i) is quite standard \cite{BraunBook}, and consists in
approximating $r_+$ and $r_-$ with suitable rational numbers
$N/N_+$ and $N/N_-$ (with $N_\pm = [N/r_\pm]$ meaning the
nearest-integer of $N/r_\pm$). The second (ii) uses, instead, a
finite $N$ and machine-precision values of $r_+$ and $r_-$.
For irrational $r_+$ and/or $r_-$, both methods introduce a finite-size error
\begin{equation} \label{Delta:eq}
\Delta= \Delta_++\Delta_- \;,
\quad {\rm with}\quad
\Delta_\pm =
2 \pi \left|N_\pm-\frac N{r_\pm}\right| \;,
\end{equation}
in the relative phase of the three periodic objects involved in the model.
In method (i) this error is distributed uniformly through the
chain, while in method (ii) it remains concentrated where the
boundary condition applies, at sites $1$ and $N$.
For a generic $N$, $\Delta$ is of order unity, but, as illustrated in
Fig.~\ref{model:fig}b (reporting the $N$ values where $\Delta$ is smaller
than for any other $N'<N$), it can be made arbitrarily small by choosing
suitably large integers $N,N_{\pm}$.
In both methods we obtain, as long as $\Delta<0.1$, fairly stable and
consistent results, which converge to the same limit for large $N,N_{\pm}$,
where $\Delta$ vanishes and full incommensurability is restored.
In a $(r_+,\ r_-\!=\!r_+^2)$ scheme, the speed of convergence depends
crucially on the nature of the incommensurability of $r_+$.
Quadratic irrational values of $r_+$ are, thanks to their {\em periodic}
continued-fraction representation \cite{Khinchin}, especially
advantageous and yield a rapidly vanishing phase error $\Delta\propto N^{-1}$
(the example quadratic irrationals of Fig.~\ref{model:fig}b include
the golden mean $\phi=(1+\sqrt{5})/2 \simeq 1.618$,
with continued-fraction coefficients $(1,1,1,...)$,
and $(2+\sqrt{2})/3 \simeq 1.138$,
with continued-fraction coefficients $(1,7,4,8,4,8,4,8,\cdots)$).
By contrast, for non-quadratic irrationals, and/or for general
$r_-\neq r_+^2$,
%
%
the phase error vanishes much more slowly $\Delta\propto N^{-1/2}$
(the examples of Fig.~\ref{model:fig}b include the spiral mean
$\sigma \sim (1,3,12,1,1,3,2,3,2,4,\cdots) \simeq 1.325$
solution of $\sigma^3=\sigma+1$,
and $\pi/2 \sim (1,1,1,3,31,1,145,1,4,2,\cdots)\simeq 1.571$).
%
Finally, for the class $(r_+,\ r_-\!=\!\theta\, r_+/[r_+-1])$, with
rational $\theta$, the error decays rapidly $\Delta\propto N^{-1}$,
regardless of the irrational nature of $r_+$.

Finite chains with open boundary conditions (OBC) were also considered,
with similar results briefly discussed below,
and dealt with in greater detail in Ref.~\cite{Cesaratto07}.
Similar models were considered in the past \cite{Urbakh,Rozman96,Zaloj98}
missing however the ingredient of incommensurability, which is important
for the novel features shown here.
A previous study of the present model \cite{VanossiPRL} achieved
the sliding through application of a driving
to one of the two substrates, via an {\it additional spring}.
That procedure obscures the quantization phenomena which are
instead uncovered when sliding occurs with a constant {\it
velocity} $v_{\rm ext}$.

Upon sliding the substrates, $v_{\rm ext} \neq 0$, the lubricant chain
slides too.  However, it generally does so in a nontrivial manner: the
time-averaged relative chain velocity $w=v_{\rm cm}/v_{\rm ext}$, is
generally {\it asymmetric}, namely different from 1/2.
Even more surprisingly, $w$ is exactly {\it quantized}, for large parameter
intervals, to plateau values that depend solely on the chosen
commensurability ratios.
The asymmetrical $w$-plateaus are generally very stable, and
insensitive to many details of the model, indicating an
intrinsically topological nature. We show that they are the
manifestation of a certain density of topological solitons ({\it
kinks}) in the lubricant versus the ``denser'' substrate potential
with periodicity $a_+$. These solitons are set into motion with
the external driving velocity $v_{\rm ext}$ by the sliding
``sparser'' substrate with periodicity $a_-$.

\section{The plateau dynamics}

\begin{figure}
\centerline{\epsfig{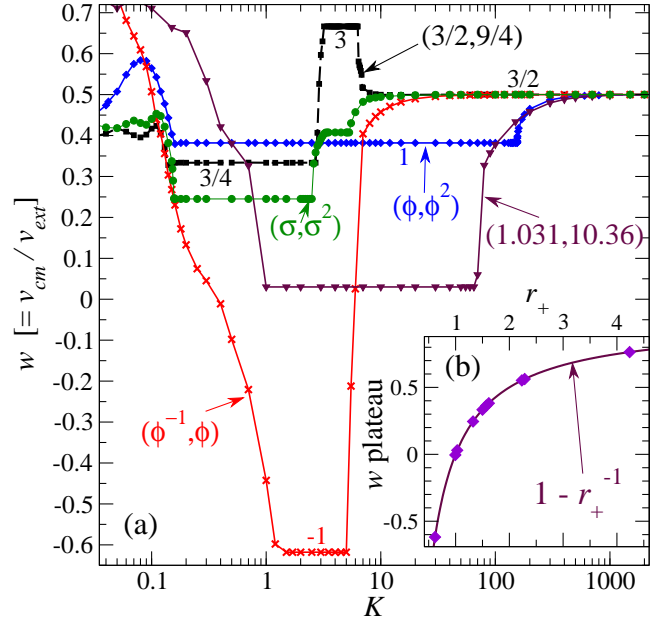}}
\caption{\label{velcm:fig}
%
%
(a) Average drift velocity ratio $w=v_{\rm cm}/v_{\rm ext}$ of the chain as
a function of its spring stiffness $K$ for different length ratios
$(r_+,r_-)$: commensurate $(3/2,9/4)$, golden mean (GM) $(\phi,\phi^2)$,
spiral mean $(\sigma,\sigma^2)$, $r_+$ close to unity
$(\frac{1031}{1000},\frac{1031}{1000}\sqrt{101})\simeq (1.031,10.36)$ --
dilute kinks, and $r_+$ smaller than unity $(\phi^{-1},\phi)$ -- antikinks.
When relevant, plateaus are labeled by the rational $\theta$ of
Eq.~\eqref{commensurate}.
All simulations are carried out in PBC with $\gamma=0.1$, $v_{\rm ext}=0.1$.
The $(\phi,\phi^2)$ plateau value is $w=0.381966\dots$, identical to
$1-\phi^{-1}$ to eight decimal places.
(b) The main plateau speed $w$ as a function of $r_+$.
}
\end{figure}

We now turn to illustrating the results.
We integrate the equations of motion (\ref{eqmotion:eqn}) starting from
fully relaxed springs ($x_i=i\, a_0$, $\dot{x}_i=v_{\rm ext}/2$), by a
standard fourth-order Runge-Kutta method.
After an initial transient, the system reaches its dynamical stationary
state, at least so long as $\gamma$ is not exactly zero.
Figure~\ref{velcm:fig}a shows the resulting time-averaged center-of-mass
(CM) velocity $v_{\rm cm}$ as a function of the chain stiffness $K$ for four
representative $(r_+,r_-)$ values:
three with $r_+>1$ and $r_-=r_+^2$, and one with $r_+<1$.
We find that $w=v_{\rm cm}/v_{\rm ext}$ is generally a complicated function of $K$,
with flat plateaus and regimes of continuous evolution, in a way which is
qualitatively similar for different cases.
The main surprise is that all plateaus show perfectly flat
$w$ values, that are exactly constant (quantized) to all figures of numerical
accuracy, the precise value strikingly independent of $K$,
of $\gamma$, of $v_{\rm ext}$, and even of $F_-/F_+$.

\begin{figure}
\centerline{
\epsfig{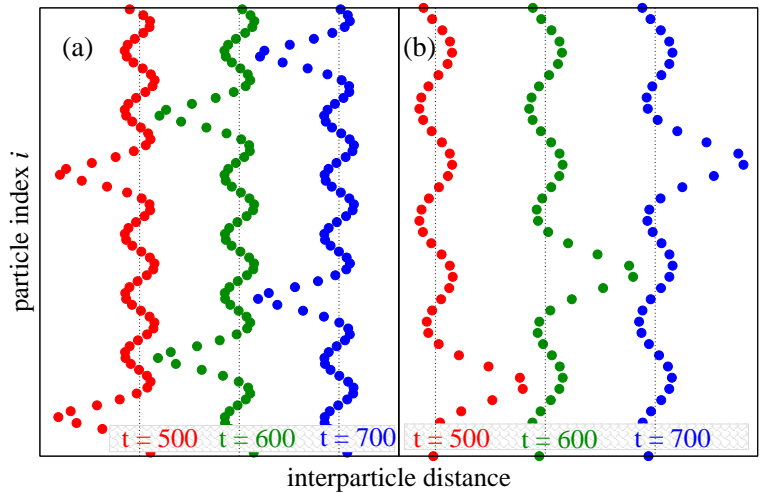}
}
\caption{\label{kinks:fig}
%
%
Snapshots of the distance between neighbor lubricant particles in the
chain $x_i-x_{i-1}$ at three successive time frames.
All simulations are carried out with $\gamma=0.1$, $v_{\rm ext}=0.1$,
$r_-\simeq 10.36$, and $K=10$ (inside the main plateau).
(a) $r_+=1.031$ (kink density $\delta/a_+= 0.031/a_+$); (b) $r_+=0.995$
(anti-kink density $|\delta|/a_+ = 0.005/a_+$).
}
\end{figure}

To explore the origin of the $w$ plateaus, we analyze the dynamics for a
large number of values of $(r_+,r_-)$, and observe that:
(i) at least one velocity plateau as a function of $K$ occurs
for a wide range of $(r_+,r_-)$;
(ii) additional narrower secondary plateaus often arise for stiffer
chain (larger $K$, see Fig.~\ref{velcm:fig}a);
(iii) the velocity ratio $w$ of the main plateau -- the first plateau
found for increasing $K$ -- satisfies
\begin{equation}
w=1-r_+^{-1}  \;,
\end{equation}
for a large range of $(r_+,r_-)$, see Fig.~\ref{velcm:fig}b.

As sketched in our recent note \cite{Vanossi06}, we can understand these results as follows.
Consider for illustration a situation of quasi-commensuration of the chain to
the $a_+$ substrate: $r_+=1+\delta$, with $\delta \ll 1$.
The slight commensurability mismatch induces a density
\begin{equation}\label{rhosol}
\rho_{sol}=\frac{\delta}{a_+}=\frac{r_+-1}{a_+}
\end{equation}
of solitons (or kinks),
essentially substrate minima holding two particles, rather than one
\cite{BraunBook}.
Assume that the second, less oscillating $a_-$ slider, which moves at velocity
$v_{\rm ext}$, will {\em drag} the kinks along: $v_{sol}=v_{\rm ext}$.
If $\rho_0 = 1/a_0 = r_+/a_+$ is the linear density of lubricant particles,
mass transport will obey $v_{\rm cm}\, \rho_0 = v_{sol}\, \rho_{sol}$.
This yields precisely
\begin{equation}
w=\frac{v_{\rm cm}}{v_{\rm ext}}=
\frac{\rho_{sol}}{\rho_0}=\frac{\delta}{r_+}=1-\frac 1{r_+}
\,.
\end{equation}
Thus the quantized velocity plateaus appear to arise because the smoother
slider (whose exact period $a_-$ is irrelevant, but is the one further away
from the chain's periodicity, $r_->r_+$) drags the lattice of kinks, of
fixed and given density, at its own full external speed $v_{\rm ext}$, as
illustrated Fig.~\ref{kinks:fig}a.
As shown in Fig.~\ref{velcm:fig}b, the physics of kink dragging is not limited
to the quasi commensurate case  $\delta \ll 1$, but extends to all values of
$|\delta|\sim 1$, where individual kinks can hardly be singled out.
Amusingly, this mechanism also works for $\delta<0$ i.e.\ $r_+<1$ . Here
kinks are replaced by anti-kinks, and we find that the chain moves in the
{\em opposite} direction to the external driving $v_{\rm ext}$
($w<0$, see Fig.~\ref{velcm:fig}a for $r_+=\phi^{-1}$). Exactly as holes
in a semiconductor, anti-kinks (carrying a negative ``charge'') dragged at
velocity $+v_{\rm ext}$ effectively produce a {\it backward} net motion
of the lubricant layer.
This ``upstream'' motion of the anti-kinks (regions of increased inter-particle
separation, and of decreased local particle density) is illustrated in Fig.~\ref{kinks:fig}b.

Before moving on, we must explain what makes the chain select the substrate
with which kinks are formed, as opposed to the other substrate merely acting as a moving
brush dragging the kinks along. As it turns out, kinks are formed relative to the
nearer-in-register substrate, which by construction has here the period $a_+$.
The density \eqref{rhosol} of kinks is such that they can be either
commensurate or incommensurate relative to the periodicity $a_-$ of the
second substrate.
The soliton lattice is commensurate to the $a_-$ substrate whenever the
ratio of $a_-$ to the average inter-kink distance $\rho_{sol}^{-1}$ is a
rational number $\theta$, i.e. when
\begin{equation}\label{commensurate}
r_-=\frac{r_+}{r_+\!-\!1}\; \theta
\,.
\end{equation}
We find that this condition, for $\theta=1$, is particularly beneficial to
formation of a large and stable main plateau.
Indeed $(r_+,r_-)=(\phi,\frac{\phi}{\phi-1})=(\phi,\phi^2)$ does satisfy
Eq.~\ref{commensurate} for $\theta=1$, and its $K$-plateau covers almost
three decades, as shown in Fig.~\ref{velcm:fig}a.
We find similarly wide plateaus for $(r_+,\frac{r_+}{r_+\!-\!1})$, with
$r_+=\sigma$, $r_+=\pi/2$ and $r_+=\pi^{1/2}$.

\begin{figure}
\centerline{
\epsfig{file=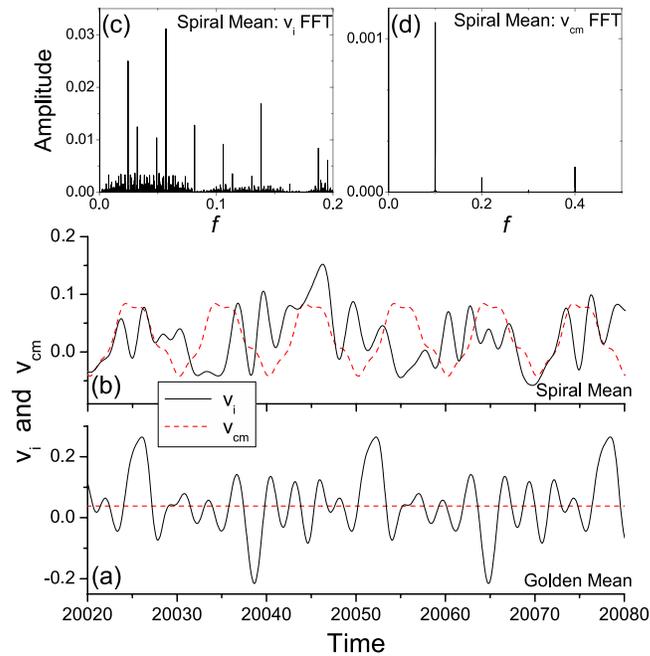,width=8.5cm,angle=0,clip=}
}
\caption{\label{evol:fig}
%
%
Time evolution of a particle velocity $\dot{x}_i$, and of the
chain CM velocity $v_{\rm cm}$ (fluctuations rescaled by a factor
50), for the GM (a) and SM (b) cases of Fig.~\ref{velcm:fig}.
Amplitudes of the Fourier spectrum of $\dot{x}_i(t)$ (c) and of
$v_{\rm cm}(t)$ (d) for $r=\sigma$. Individual particle spectra
have identical amplitudes, and differ only in the phases, which
leads to a remarkable cancellation in the $v_{\rm cm}$ power
spectrum. Here $K=1$, $\gamma=0.1$, and $v_{\rm ext}=0.1$. }
\end{figure}

\subsection{Individual particle motion and CM periodic dynamics}

The motion of the individual particles is also very instructive.
Figure~\ref{evol:fig} shows the time evolution of the velocity of a single
chain particle $\dot{x}_i$, and of $v_{\rm cm}$, for a value of $K$ inside a
plateau, $K=1$, for $r_+=\phi$ and $r_+=\sigma$ of Fig.~\ref{velcm:fig}.
Here two clear kinds of behavior emerge.
The single-particle motion of each particle in the chain is, during
motion in the GM plateau, perfectly time-periodic.
A similar periodic dynamics is found, in appropriate regimes, for all
rational and {\em quadratic} irrational $(r_+,\ r_-\!=\!r_+^2)$ values
which we tested.
%
Similarly, periodic motion emerges on the plateaus of $(r_+,r_-)$ satisfying
Eq.~\eqref{commensurate}, for arbitrary $r_+$.
On the contrary, single-particle motion in the $(\sigma,\sigma^2)$ spiral
mean (SM) case is definitely not periodic.
The Fourier spectrum of the particle motion, shown in Fig.~\ref{evol:fig}c,
confirms that SM sliding yields only quasi-periodic orbits with two prominent
incommensurate frequencies $f_+$ and $f_-$, which can be interpreted as the
average frequency of encounter of a generic particle with the two substrate
periodic corrugations, $f_+= v_{\rm cm}/a_+$ and
$f_-= (v_{\rm ext}-v_{\rm cm})/a_-$ respectively.
There is however a complete and exact phase cancellation between
the Fourier spectra of different
chain particles (all having the same amplitude spectrum, with different
phases), giving rise to a strictly periodic motion of the chain CM even in
the SM case.
Periodic CM velocity oscillations around an exactly quantized drift
velocity is a common and generic feature of all plateaus in the chain
dynamics.
These periodic oscillations can be understood as the solitons moving at
velocity $v_{\rm ext}$ encountering the periodic Peierls-Nabarro potential
\cite{BraunBook} of period $a_+$.
The corresponding frequency of encounter $v_{\rm ext}/a_+$ is clearly visible
in Fig.~\ref{evol:fig}d.
We observe that the Peierls-Nabarro barrier vanishes (yielding a strictly
constant $v_{\rm cm}$) for all cases where the particle motion is periodic,
particularly all $(r_+,r_-)$ satisfying Eq.~\eqref{commensurate}.
When such a periodic single-particle motion occurs for a $a_-$-commensurate
soliton lattice, Eq.~\eqref{commensurate}, the frequency ratio
\begin{equation}
\frac{f_+}{f_-}= r_- \frac{(r_+\!-\!1)}{r_+} = \theta \,.
\end{equation}
%
As this condition is equivalent to Eq.~\eqref{commensurate}, this analysis
attributes to the rational $\theta$ the double significance of (i) the coverage
fraction of solitons on the substrate of periodicity $a_-$, and (ii) the
ratio between the average frequencies $f_+$ and $f_-$ of encounter of the
features of the two substrates.

Low driving velocities $v_{\rm ext}$ are beneficial to the appearance and to the
width of velocity plateaus.
For increasing $v_{\rm ext}$, the plateaus shrink and eventually disappear,
still remaining exact while they do so.
The critical $v_{\rm ext}$ where the plateaus eventually disappear depends
on $K$, but is usually smaller than unity, for the parameters of
Fig.~\ref{velcm:fig}.
Fully commensurate $\theta=1$ plateaus such as that of the $(\phi,\phi^2)$
case are especially wide and robust against an increase of $v_{\rm ext}$
(for the parameters of Fig.~\ref{velcm:fig} and $K=4$, up to $v_{\rm
ext}\simeq 1.5$) and other perturbations.
Within the class of $(r_+,\,r_+^2)$ length ratios, $r_+=\phi$ appears
therefore, in the present dynamical context, as the ``most {\em commensurate}''
irrational, at variance with static pinning in the standard
FK model, where the opposite occurs \cite{BraunBook,Aubry83}.

All the above is for PBC. As anticipated, OBC simulations
\cite{Cesaratto07} show however that the PBC used are not crucial
to the plateau quantization, which occurs even for a lubricant ``patch'' of
finite and not particularly large size $N$, perhaps such as a hydrocarbon
chain molecule would be in 1D, or an interposed graphite flake in
two-dimensions (2D).
The finite lubricant size does not remove the plateaus. On the contrary, it
permits and favors the development of robust velocity plateaus as a
function of stiffness. The plateau values are not identical to those of the
infinite chain with PBCs. The reason for that is that finite size allows
for an overall chain-length re-adjustment which spontaneously promotes
single-particle {\em periodic} oscillations, effectively satisfying
Eq.~\eqref{commensurate}.

\begin{figure}
\centerline{
\epsfig{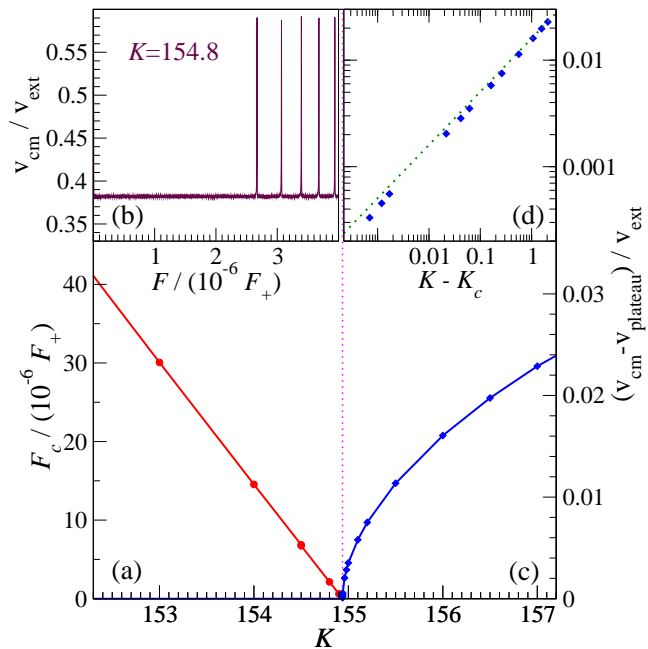}
}
\caption{\label{critical:fig}
%
Dynamical depinning force $F_c$ for the GM plateau (a), extracted by a slow
adiabatic increase of the force $F$ applied to all particles, until
intermittencies appear, signaling collective slips (b). The critical
behavior of the average velocity for $K\to K_c$ (c,d), showing the typical
square-root singularity (dotted line) associated to type-I intermittencies.
Here $\gamma=0.1$, and $v_{\rm ext}=0.1$ were used.
}
\end{figure}

\subsection{Dynamical incompressibility and tribological dissipation}

Returning to the general infinite chain case, the finding of exact plateaus
implies a kind of ``dynamical incompressibility'', namely identically
null response to perturbations or fluctuations trying to deflect the
CM velocity away from its quantized value.
In order to probe the robustness of the plateau attractors, we can introduce
an additional constant force $F$ acting on all particles in the chain,
whose action is to try to alter the force-free sliding velocity away from its
plateau value.
As expected, as long as $F$ remains sufficiently small, it does perturb the
single-particle motions but has no effect whatsoever on $w$, which remains
exactly pinned to the attractor ($v_{\rm cm}\equiv v_{\rm plateau}$).
The plateau dynamics is only abandoned above a critical force $F_c$.
This dynamical depinning takes place through a series of type-I
intermittencies \cite{Berge84}, as shown in Fig.~\ref{critical:fig}b
where $v_{\rm cm}(t)$ is plotted against a slow adiabatic ramping of $F$.

A precise value of $F_c$ can be obtained by ramping $F$ with time with a
gentle enough rate of increase or, alternatively, by a Floquet-Lyapunov
linear stability analysis.
This procedure is standard \cite{Jose98} and amounts to studying the
eigenvalues of a $2N\times 2N$ matrix ${\bf R}$ whose columns are the
linearized time-evolutions, over one period $T$, of the standard basis
vectors ${\bf e}_i=\delta_{i,j}$ in phase-space.
The motion is stable if the (complex) eigenvalues $\nu_i$ of ${\bf R}$ are
smaller than unity, $|\nu_i|<1$; intermittencies of type I arise when the
largest (in modulus) eigenvalue of ${\bf R}$, $\nu_{\rm max}$, reaches the
point $1$ in the complex plane, $\nu_{\rm max}\to 1$, which is indeed what we
find at the edge of the GM plateau.
The value of $F_c$ is a function of the parameters, and $F_c$ vanishes
linearly when $K$ approaches the border $K_c$ of the plateau, as in
Fig.~\ref{critical:fig}a.
The depinning transition line $F_c$ shows a jump $\Delta v$ in the average
$v_{\rm cm}$ and a clear hysteretic behavior \cite{Vanossi07Hyst} as $F$
crosses $F_c$, for not too large values of $K<K_c$.
As can be expected, $\Delta v$ decreases to $0$ as $K$ increases.
Around $K=K_c$ the depinning transition is continuous.  The specific point
$K=K_c$ represents the zero-$F$ crossing of the depinning line, which
extends to negative $F$ above $K_c$.
The precise value of $K_c$ depends on parameters such as $v_{\rm ext}$ and
$\gamma$.
As $K$ approaches $K_c$ from above (no external force), $v_{\rm cm}$ approaches
$v_{\rm plateau}$ in a critical manner, as suggested in
Fig.~\ref{velcm:fig}(a).
This is detailed in Fig.~\ref{critical:fig}(c,d), where the critical behavior
is shown to be $\Delta v\propto (K-K_c)^{1/2}$, the value typical of
intermittencies of type I \cite{Berge84}.
For $K > K_c$, in fact the chain spends most of its time moving
at $v_{\rm cm}(t) \simeq v_{\rm plateau}$, except for short bursts at
regular time-intervals
$\tau$, where the system as a whole jumps ahead by $a_0$, i.e.\ an extra
chain lattice spacing (collective slip).
The characteristic time $\tau$ between successive collective slips
diverges as $\tau\propto(K-K_c)^{-1/2}$ for $K\to K_c$, consistent with
the critical behavior of $w$.
We verified that the $w$-plateaus for more general values of $r_+$ and of
$r_-$ show the same kind of infinite stiffness, and a critical decrease of
$F_c$ near the plateau edge, similar to that of Fig.~\ref{critical:fig}a
for the GM.

\begin{figure}
\centerline{
\epsfig{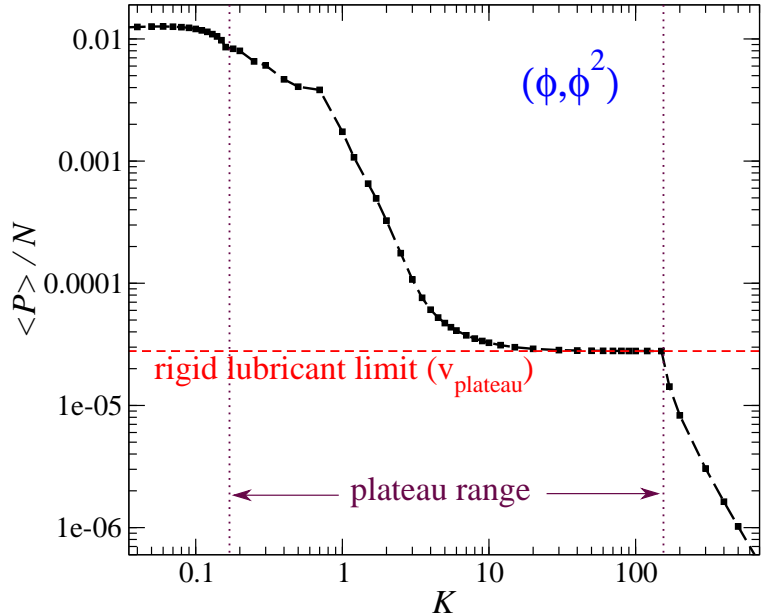}
}
\caption{\label{dissipation:fig}
%
The average dissipated power per particle for the GM
$(r_+,r_-)=(\phi,\phi^2)$ model as a function of the spring stiffness $K$.
All dynamical parameters are as in Fig.~\ref{velcm:fig}.
%
Two dotted lines indicate the plateau range. A dashed horizontal line marks
the theoretical lowest dissipation limit of a rigid lubricant in a purely
translational motion at speed $v_{\rm plateau}=(1-\phi^{-1})\,v_{\rm ext}$.
}
\end{figure}

We have investigated the {\it tribological} properties of the model
in the plateau regime.
As the dissipative term makes no work on particles that move at velocity
$\dot{x}_i= v_{\rm ext}/2$, it is particularly convenient to evaluate the
dissipated power in the ``symmetric'' frame of reference where the upper
chain moves at velocity $\frac{v_{\rm ext}}2$ and the lower chain moves at
velocity $-\frac{v_{\rm ext}}2$.
In this frame of reference, the instantaneous total power
dissipated by the $\gamma$ term amounts to
\begin{eqnarray}
P = \sum_i \dot{x}_i \cdot
\left( 2 \, \gamma \, \dot{x}_i \right)
=2\,\gamma\,\sum_i\dot{x}_i^2
=\frac{4\,\gamma}{m} \, E_{\rm kin} \;,
\end{eqnarray}
where $E_{\rm kin}$ is the total kinetic energy of the particles in the
symmetric frame of reference.
This power fluctuates, but eventually, in the dynamical stationary state
its average value $\langle P\rangle$ equals on average the mean power
necessary to maintain the motion of the sliders \cite{VanossiPRL}.
Figure~\ref{dissipation:fig} shows the dissipated power per particle in a
wide range of $K$ values for the $(\phi,\phi^2)$ model.
It is apparent that the friction decreases as the springs become stiffer
and stiffer, where fluctuations are of decreasing amplitude.
The plateau region does not appear to be especially conspicuous in this
frictional decrease, except for a net drop at the end of the plateau, due to the
lubricant moving closer and closer to the symmetric velocity $v_{\rm ext}/2$.

\section{Conclusions}

The phenomena just described for a model 1D system might have 2D
counterparts, potentially observable in real systems.
Nested carbon nanotubes \cite{Zhang}, or confined one-dimensional
nanomechanical systems \cite{Toudic_06}, are one of several possible arenas
for the phenomena described in this paper.
Though speculative at this stage, an obvious question is what aspects of
the phenomenology just described might survive in 2D, where tribological
realizations, such as the sliding of two hard crystalline faces with, e.g.,
an interposed graphite flake, are conceivable.
Our results suggests that the lattice of discommensurations -- a Moir\'e
pattern-- formed by the flake on a substrate, could be dragged by the other
sliding crystal face, in such a manner that the speed of the flake as a
whole would be smaller, and quantized. This would amount to the slider
``ironing'' the solitons onward.
Dienwiebel {\it et al.}\ \cite{Dienwiebel04} demonstrated how
incommensurability may lead to virtually friction-free sliding in such a
case, but no measure was obtained for the flake relative sliding velocity.
%
Unlike our model, real substrates are not rigid nor ideal, but subject to
thermal expansion and characterized by defects.  Nevertheless the ubiquity
of plateaus shown in Fig.~\ref{velcm:fig}, and their topological origin,
suggests that these effects would not remove the phenomenon.
A real-life situation with a distribution of differently oriented
crystalline micro-grains, each possessing a different incommensurability,
is also potentially interesting; each grain, we expect, will tend to
stabilize a certain average CM velocity depending on its
incommensurability.
Other realizations or applications inspired by the physics described by our
model might be accessible, notably in grain-boundary motion, in the sliding
of optical lattices \cite{optical} or of charge-density-wave systems
\cite{Gruener_cdw}.

\section*{Acknowledgments}
We are grateful to O.M. Braun for invaluable discussions.
This research was partially supported by PRRIITT (Regione
Emilia Romagna), Net-Lab ``Surfaces \& Coatings for Advanced Mechanics and
Nanomechanics'' (SUP\&RMAN) and by MIUR Cofin 2004023199, FIRB RBAU017S8R,
and RBAU01LX5H.

\section*{References}


\end{document}